# Bionic Reflex Control Strategy for Robotic Finger with Kinematic Constraints

**Narkhede Kunal Sanjay**
*Department of Mechanical Engineering*
*Indian Institute of Technology Kharagpur*
*Kharagpur-721302,West Bengal, India*
kunalnarkhede@iitkgp.ac.in

**Shyamanta M. Hazarika**
*Department of Mechanical Engineering*
*Indian Institute of Technology Guwahati*
*Guwahati-781039, Assam, India*
s.m.hazarika@iitg.ac.in

*Abstract*---This paper presents a bionic reflex control strategy for a kinematically constrained robotic finger. Here, the bionic reflex is achieved through a force tracking impedance control strategy. The dynamic model of the finger is reduced subject to kinematic constraints. Thereafter, an impedance control strategy that allows exact tracking of forces is discussed. Simulation results for a single finger holding a rectangular object against a flat surface are presented. Bionic reflex response time is of the order of milliseconds.
*Keywords*---bionic reflex control, impedance control

## I. INTRODUCTION

Human brain together with the large number of sensors on the hand ensures stable grasps even under slippage. This is achieved by rapid adjustment of the grasping force in response to static information that includes object mass, stiffness, and friction, However even the most advanced of the prosthetic hand lacks this basic reflex. Only recently there has been work on bionic reflex control system designed to reconstruct the humanoid reflex control function for a prosthetic hand [1]. In bionic reflex, possibly our most common reaction is to increase the grasp forces subject to stiffness of the object being grasped. Therefore, tracking force holds promise for bionic reflex. Impedance based force tracking under unknown environment for robotic manipulator is well established [2,3,4]. Force tracking for grasp control of prosthetic hands have been explored [5] including control of slippage [6].

In this paper we present a bionic reflex control strategy for a kinematically constrained robotic finger. The second section presents the dynamics of kinematically constrained robotic finger. The dynamic reduction is done closely following [7]. In the third section we have explained in details the impedance based force control methodology and our proposed bionic reflex control strategy. In the fourth section we discuss the simulation results concluding the paper in section five.

## II. DYNAMICS OF KINEMATICALLY CONSTRAINED FINGER

### A. Dynamic model of a finger

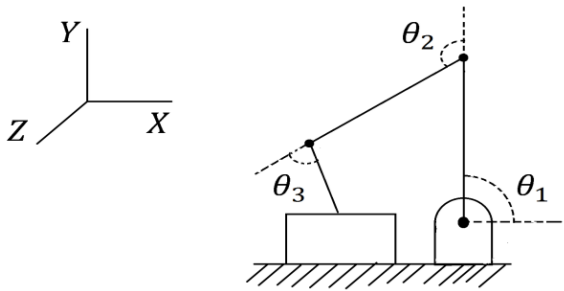

Fig.1 : Schematic of the finger holding an object

Fig. 1 shows the schematic diagram of a single finger. The finger is holding an object against the palm. The finger consists of three joints. Even though each joint could be seen as independent, for natural curling the joints have a constrained specified which leads to reduction in independent DoFs. This is done closely on lines of the discussion on model reduction in [7].

First, let us consider the dynamic equations of **3** degrees of freedom planer finger subjected to **2** holonomic kinematic constraints. These equations are based on the Euler-Lagrangian formulation and can be represented as:

$$D(\theta)\ddot{\theta} + C(\theta,\dot{\theta})\dot{\theta} + G(\theta) + \tau_e = \tau + A^T(\theta)\lambda \quad (1)$$

Where $\theta \in \mathbb{R}^3$ is the angular position vector; as a convention bold face is used for vectors and matrices. $\tau \in \mathbb{R}^3$ is the generalized torque vector. $D(\theta) \in \mathbb{R}^{3*3}$ is symmetric and positive definite inertia matrix. $C(\theta,\dot{\theta}) \in \mathbb{R}^{3*3}$ is the the centripetal and Coriolis torque matrix. $G(\theta) \in \mathbb{R}^3$ is the gravitational torque vector. $\tau_e = J^T(\theta)f_e$ is the external contact torque acting on the end effector due to external contact force $f_e \in \mathbb{R}^3$. $A^T(\theta)\lambda$ is the constraint force term, where $\lambda$ is the Lagrange multiplier and $J(\theta) \in \mathbb{R}^{2*3}$ is the finger Jacobian matrix. For further details, please refer to [8].

### B. Model Reduction

The finger is subjected to kinematic constraints defined by

$$\theta_1 - \left(\frac{4}{5}\right) * \theta_2 = 0 \quad (2)$$

$$\theta_1 - \left(\frac{6}{5}\right) * \theta_3 = 0 \quad (3)$$

Where $\theta_1$, $\theta_2$ and $\theta_3$ are the joint angles of proximal, middle and distal phalanxes respectively. Differentiating the above equations with time we get

$$A(\theta)\dot{\theta} = 0 \quad (4)$$

Where $A(\theta) = \begin{bmatrix} 1 & -\frac{4}{5} & 0 \\ 1 & 0 & -\frac{6}{5} \end{bmatrix}$ and $\theta = \begin{bmatrix} \theta_1 \\ \theta_2 \\ \theta_3 \end{bmatrix}$.

From equation (2) and (3) we can write

$$\boldsymbol{\theta} = \begin{bmatrix} \frac{1}{5} \\ \frac{4}{5} \\ \frac{5}{6} \end{bmatrix} \theta_1 = \boldsymbol{L}\theta_1. \tag{5}$$

As $\boldsymbol{L}$ lies in the null space of $\boldsymbol{A}$, the following equation is satisfied

$$\boldsymbol{AL} = \boldsymbol{L}^T \boldsymbol{A}^T = \boldsymbol{0}.$$

Also differentiating equation (5) we get

$$\ddot{\boldsymbol{\theta}} = \boldsymbol{L}\ddot{\theta}_1. \tag{6}$$

From equation (1) and (6) the reduced dynamic equation of the system can be written as

$$D'(\theta_1)\ddot{\theta}_1 + C'(\theta_1, \dot{\theta}_1)\dot{\theta}_1 + G'(\theta_1) + \boldsymbol{L}^T \boldsymbol{\tau}_e \tag{7}$$
$$= \boldsymbol{L}^T \boldsymbol{\tau}$$

Where

$$D' = \boldsymbol{L}^T D \boldsymbol{L}$$
$$C' = \boldsymbol{L}^T C \boldsymbol{L}$$
$$G' = \boldsymbol{L}^T G.$$

It should be noted that the above equation is a scalar equation.

III. CONTROL SCHEME

The reduced dynamic equation of the finger is given by

$$D'(\theta_1)\ddot{\theta}_1 + C'(\theta_1, \dot{\theta}_1)\dot{\theta}_1 + \boldsymbol{L}^T \boldsymbol{\tau}_e = \boldsymbol{L}^T \boldsymbol{\tau} = \tau' \tag{8}$$

Since the finger is in X-Y plane and the gravity acts in negative Z-direction, we have neglected the gravity term. Here $\boldsymbol{\tau}$ is defined as $\boldsymbol{\tau} = \boldsymbol{r}\tau_a$

Where $\boldsymbol{r} \in \mathbb{R}^3$ is constant actuating coefficient vector and $\tau_a$ is the actuating torque. Also $\boldsymbol{L}^T \boldsymbol{\tau}_e$ can be written as

$$\boldsymbol{L}^T \boldsymbol{\tau}_e = \boldsymbol{L}^T \boldsymbol{J}^T \boldsymbol{f}_e = H\|\boldsymbol{f}_e\| = \tau_c$$

Where $\|\boldsymbol{f}_e\|$ is the Euclidean norm of $\boldsymbol{f}_e$ which is the magnitude of normal force acting on the end effector and the object and $H$ is a scalar defined by

$$H = \boldsymbol{L}^T \boldsymbol{J}^T \boldsymbol{p} \tag{9}$$

Where $\boldsymbol{p} \in \mathbb{R}^2$ is the vector describing orientation of $\boldsymbol{f}_e$. Since this is a single degree of freedom system, if we could control the external torque acting on the system, we could control the external contact force.

A. *Force/Torque Tracking Impedance Controller*

Here we have adapted the force/torque tracking impedance controller as proposed by Jung et al. [2] to achieve asymptotic force/torque tracking.

The dynamic equation is given by

$$D'(\theta_1)\ddot{\theta}_1 + C'(\theta_1, \dot{\theta}_1)\dot{\theta}_1 + \tau_c = \tau'$$

Where $\tau_c$ is the external torque acting on the finger as it comes in contact with the object. It can be measured by either a torque sensor mounted on first joint or by a force sensor mounted on the end effector. Our objective here is to control both position and external force. The control law for impedance control is given by

$$\tau' = D'(\theta_1)U + C'(\theta_1, \dot{\theta}_1)\dot{\theta}_1 + \tau_c. \tag{10}$$

The control input $U$ is given by

$$U = \ddot{\theta}_{1d} + \left(\frac{1}{M}\right)\left(B(\dot{\theta}_{1d} - \dot{\theta}_1) + K(\theta_{1d} - \theta) + \tau_d - \tau_c\right).$$

Where $M$ is the desired inertia, $B$ is the desired damping; $K$ is the desired stiffness and $\tau_d$ is the desired contact torque/force. For force/torque tracking we replace $\theta_{1d}$ with $\theta_{1e}$, here $\theta_{1e}$ is the position of the object. For a review on impedance control, please see [9].

Substituting the control law back into the dynamic equation we get

$$M(\ddot{\theta}_{1e} - \ddot{\theta}_1) + B(\dot{\theta}_{1e} - \dot{\theta}_1) + K(\theta_{1e} - \theta) \tag{11}$$
$$= \tau_c - \tau_d.$$

Let $e = \theta_{1e} - \theta$ be the error in measurement of position. Then the above equation becomes

$$M\ddot{e} + B\dot{e} + Ke = \tau_c - \tau_d. \tag{12}$$

Here we have used two phase control law as proposed by Jung et al. [2]. The first phase is free space control in which the external contact torque/force is zero. During first phase if we set the desired torque to zero, the equation becomes

$$M\ddot{e} + B\dot{e} + Ke = 0 \tag{13}$$

basically reducing to position control. The above equation is asymptotically stable and the position error reduces to zero asymptotically and at steady state $\theta_1 = \theta_{1e}$.

In second phase that is the contact phase we set the desired stiffness $K$ to zero; the equation becomes

$$M\ddot{e} + B\dot{e} - \tau_c = -\tau_d. \tag{14}$$

For steady state, as derivatives vanishes
$$\tau_c = \tau_d \text{ or } H\|\boldsymbol{f}_e\| = H\|\boldsymbol{f}_d\|$$
where $\boldsymbol{f}_d$ is the desired force; thus enabling us to track force. In this control strategy we assume that the location of the object is known to us. This control law is quite robust and tracks force/torque irrespective of the object stiffness knowledge.

B. *Bionic Reflex Control*

In this section we propose a control method to regulate the desired force $\boldsymbol{f}_d$ which in turn will reduce the object's slip velocity to zero in real time. We make use of the robust and excellent force tracking performance of the impedance controller discussed in previous section. The desired contact force which controls the slip velocity of the object is modified. The mass and coefficient of friction of the object are unknown. We use scalar adaptive law [10] to estimate these parameters. The dynamics of the object falling down in presence of gravity and contact friction force assuming Coulomb friction model is given by

$$m'\dot{v} = m'g - \mu f_e \qquad (15)$$

Where $m'$ is the mass of the object; $\mu$ is the coefficient of friction and $f_e$ is the norm of contact force between object and the finger. The equation can also be written as

$$m\dot{v} = mg - f_e \qquad (16)$$

Where $m$ is $m'/\mu$. Let $\dot{v} - g = u$. So the equation can be written as

$$u = -f_e/m = af_e. \qquad (17)$$

Here $a = -\frac{1}{m} = -\mu/m'$ is the unknown parameter to be estimated in order to control $v$.

We define $\hat{a}$ as an estimate of $a$ and generate the predicted value of $\hat{u}$ of the output $u(t)$ as

$$\hat{u} = \hat{a} f_e \qquad (18)$$

The error in predicted and original value is defined as

$$\epsilon = u - \hat{u} = (a - \hat{a}) f_e = \tilde{a} f_e \qquad (19)$$

Now to minimize this error we use gradient or Newton's method. By applying this method we obtain the update law for $\hat{a}$ as $\dot{\hat{a}} = \alpha \epsilon f_e$, $\hat{a}(0) = a_0$. Where $\alpha$ is the adaptive gain and $a_0$ is the initial guess value.

Now in order to make the slip velocity zero, we define the control input $f_e$ as

$$f_e = \hat{m}(g + bv) = (-1/\hat{a})(g + bv) \qquad (20)$$

Substituting this control law in the dynamic equation we get

$$m\dot{v} = mg - \hat{m}g - \hat{m}bv$$

At steady state $\hat{m} = m$. So equation becomes

$$\dot{v} = -bv$$
$$v(t) = v(0)\exp(-bt) \qquad (21)$$

Hence by adjusting the value of $b$ we can ensure that the slip velocity of the object becomes zero in real time. Now to achieve near real time object gripping we implement this control law as

$$||f_d|| = \hat{m}(g + bv) \qquad (22)$$
$$\tau_d = H||f_d|| = H\hat{m}(g + bv). \qquad (23)$$

Here we modify the desired force/torque based on the above control law.

Due to the impedance controller, the contact force between the object and the finger tracks the desired force in real time and this force regulates the velocity of the object. Therefore our proposed bionic reflex control strategy is simple, easy to implement, robust and does not require knowledge of object mass and coefficient of friction between finger and object. Fig. 2 shows the schematic diagram of the bionic reflex control strategy.

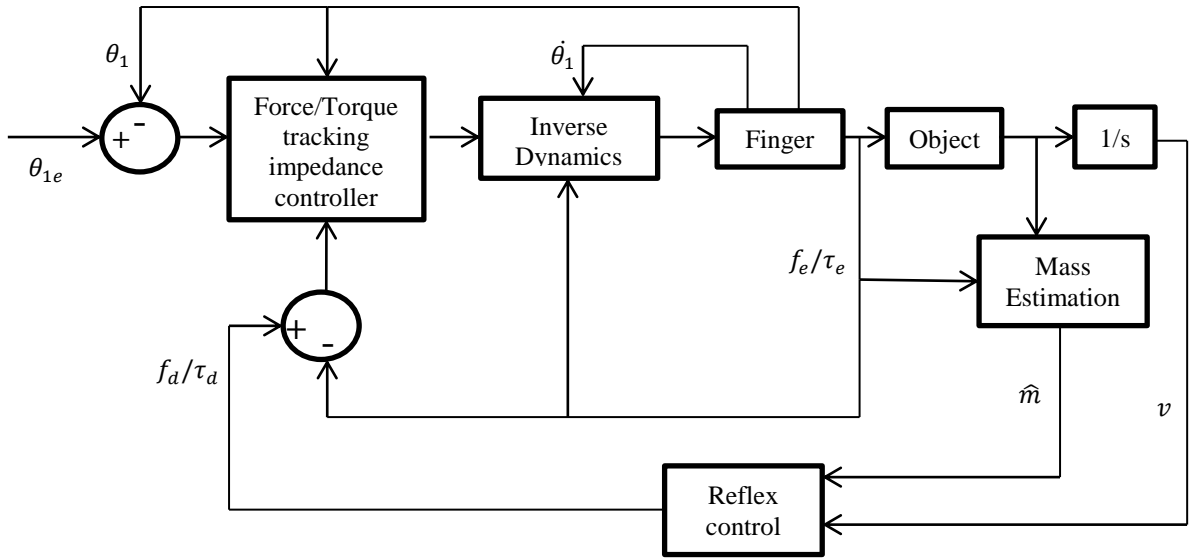

Fig.2 : Schematic Diagram of Bionic Reflex Control Strategy

## IV. SIMULATION RESULTS AND DISCUSSION

### A. Angular Position tracking

The efficiency and effectiveness of our control strategy is verified with numerical simulations. The simulations were performed in Simulink for different types of scenarios on a three link robotic finger. The lengths of the proximal, middle and distal phalanxes are 40 mm, 30 mm and 20 mm respectively. The masses of proximal, middle and distal phalanxes are 6.9580 g, 5.2185 g and 3.4790 g respectively. First, the free space control law is tested on the finger. A step input is given and the resulting response is recorded. Since we are giving a step input, in order to avoid instability we have neglected the inertia term. The desired damping is kept at 190 and the desired stiffness is kept at 9025. Fig. 3 shows the response of proximal phalange angle $\theta_1$. As seen from the figure, the position tracking is in near real time.

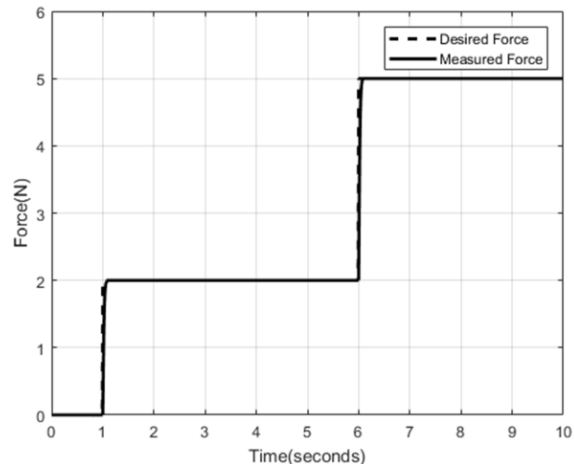

Fig. 4. Desired Force Tracking.

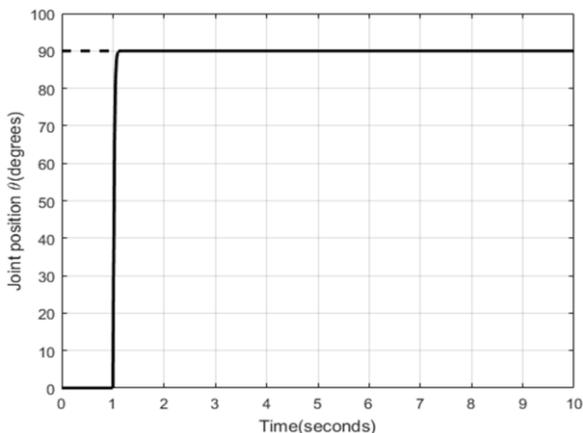

Fig. 3. Step response of joint1.

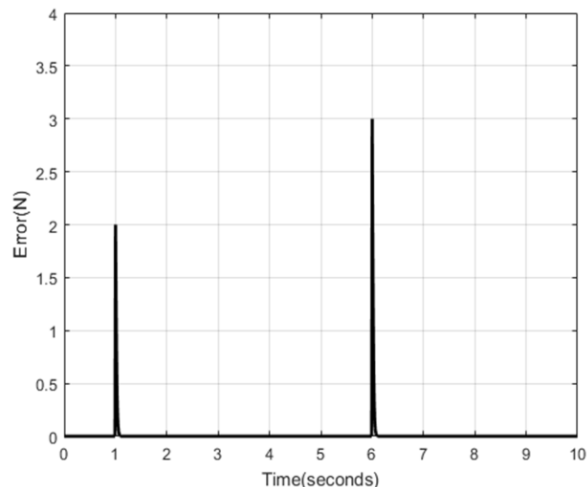

Fig. 5. Force tracking error.

### B. Desired force tracking

In this section, the contact space control law is tested on the finger. We have assumed a standard spring based contact model between object and the finger with the spring stiffness assumed to be 10000 N/m. We have given a force signal at the input which is zero till time 1 s and jumps to 2 N at 1 s, stays at 2 N till 6 s and again jumps at 5 N at 6 s. This type of force signal is applied in order to check the robustness of the controller. The impedance parameters are same as mentioned in section A. Fig. 4 shows the force tracking response of the finger as it comes in contact with the object. As seen from the figure the force tracking results are good. They almost follow the input signal profile. Here we have assumed that the object stiffness is not known. This shows that the controller can perform quite well even in unknown environment. Fig. 5 shows the force tracking error. As observed from the figure the tracking error quickly settles down to zero.

### C. Object slip velocity response

In this section, the results of bionic reflex control strategy are discussed. The finger is first allowed to grasp the object. The object is supported against the palm. At first the force is enough to hold the object. The mass/(friction coefficient) value of the object is 50 g. Now to induce slip the mass/(friction coefficient) value of the object is changed suddenly to 70 g at 5 s and again to 80 g at 10 s and the slip velocity is recorded. The value of $b$ is taken to be 20. Here we assume that the object velocity can be measured with the help of some velocity sensor [11]. The impedance parameters are same as mentioned in the previous sections. The adaptation gain is set to 320. Fig. 6 shows the slip velocity of the object. Note that complete modeling of slip is avoided by using slip velocity as a quantitative measure of slip occurring in a grasp (if any). As seen from the figure, the velocity becomes zero in near real time. Fig. 7 shows the mass/(friction coefficient) value estimation curve. According to the curve the estimated values are almost equal to the original one. The response curves clearly show that the bionic reflex control strategy is quite robust and achieves the results in near real time.

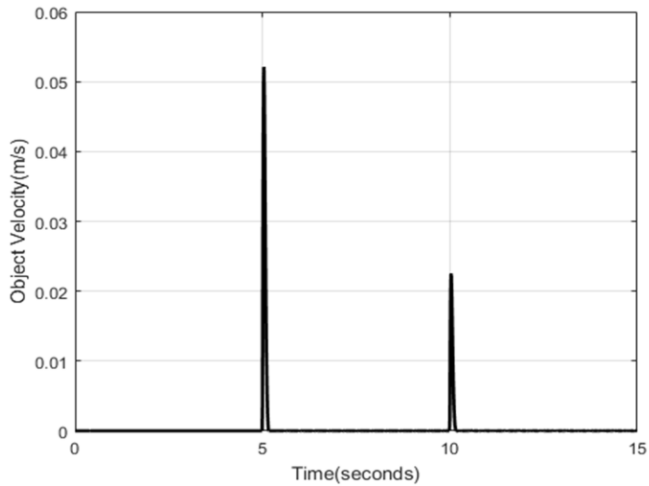
Fig. 6. Slip velocity of object.

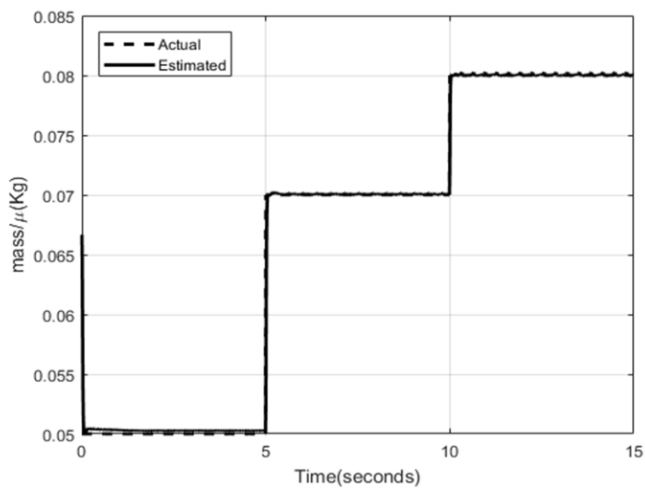
Fig. 7. Mass/(coefficient of friction) Estimation.

## V. CONCLUSION

This paper has presented a strategy that allows bionic reflex response for a single finger under kinematic constraints. As can be seen from the simulation results we have near real time response. However, this was for a single finger restricted to kinematic constraints holding a rectangular object against a flat surface. Force tracking through an impedance based controller implemented a bionic reflex strategy that ignored stiffness while updating the desired force. Consideration of stiffness within the bionic reflex strategy is part of ongoing research. Further, it would be interesting to apply the strategy to objects with different geometry.